\documentclass{Resources/UoBLab1}
\pubyear{2018}
\subjectarea{The University of Helsinki Technical Report}

\usepackage[linesnumbered,algoruled,boxed,lined]{algorithm2e}
\usepackage{fixltx2e}
\usepackage{xcolor}

\def\SP#1{\textsuperscript{\textcolor{red}{#1}}}

\begin{document}
\firstpage{1}

\title{$\Pi$-cyc: A Reference-free SNP Discovery Application using Parallel Graph Search}
\author{Reda Younsi\SP{1*}, Jing Tang\SP{2} and Liisa Holm\SP{1}}
\course{1* The Institute of Biotechnology (BI)}
\school{2  The Institute of Molecular Medicine Finland (FIMM)}
\date{September 2018}

\twocolumn[
\begin{@twocolumnfalse}
\maketitle

\begin{abstract}

\textbf{Motivation:} 
Working with a large number of genomes simultaneously is of great interest in genetic population and comparative genomics research.  Bubbles discovery in multi-genomes coloured de bruijn graph for de novo genome assembly is a problem that can be translated to cycles enumeration in graph theory. Cycle enumerations algorithms in big and complex de Bruijn graphs are time consuming.  Specialised fast algorithms for efficient bubble search are needed for coloured de bruijn graph variant calling applications. In coloured de Bruijn graphs, bubble paths coverages are used in downstream variants calling analysis.\\
\textbf{Results:} In this paper, we introduce a fast parallel graph search for different K-mer cycle sizes. Coloured path coverages are used for SNP prediction. The graph search method uses a combined multi-node and multi-core design to speeds up cycles enumeration.  The search algorithm uses an index extracted from the raw assembly of a coloured de Bruijn graph stored in a hash table.  The index is distributed across different CPU-cores, in a shared memory HPC compute node, to build undirected subgraphs then search independently and simultaneously specific cycle sizes. This same index can also be split between several HPC compute nodes to take advantage of as many CPU-cores available to the user. The local neighbourhood parallel search approach reduces the graph's complexity and facilitate cycles search of a multi-colour de Bruijn graph.  The search algorithm is incorporated into $\Pi$-cyc application and tested on a number of \textit{Schizosaccharomyces Pombe} genomes.\\

\textbf{Availability:}  $\Pi$-cyc is an open-source software available at www.github.com/2kplus2P\\

\textbf{Contact:} * reda.younsi@helsinki.fi
\end{abstract}
\end{@twocolumnfalse}
]
\section{Introduction}

De novo de Bruijn graph assemblers for short reads are mainly divided into Eulerian and Hamiltonian approaches~\cite{ga}. In both approaches, cycles, called bubbles when searching directly a de Bruijn graph, represent errors, heterozygous variants or repeats and are mainly graph artifacts that require genomic corrections for accurate assembly.  Taking a different approach, \cite{cortex_2012} showed, by using the concept of colours to represent multiple genomes, can generate multicolor graph paths, in form of bubbles, which may contain genomic variants. This coloured bubble search was first used successfully in Cortex\_var application to call genomic variants, mainly SNPs and short indels, using de Bruijn graphs \cite{cortex_2012}. Several methods using coloured bubble concept have since been developed for variant calling using de bruijn graphs~\cite{younsi2015using,Bateman}.

In this paper, we propose a fully integrated approach to SNP calling using a parallel cycle enumeration algorithm in a transformed colored de Bruijn graph (cdbg) which we call $\Pi$-cyc.  The proposed method can use a single, or several compute, nodes with multi-cores CPUs to build undirected subgraphs and search for cycles in parallel.  A similar algorithm was first used sequentially to enumerate specific $2k+2$ bubbles, k being the k-mer size used for the cdbg~\cite{younsi2015using}.  These cycles where shown to harbour a large number of real SNPs but missed a substantial number of other SNPs found in different cycle sizes, those that are normally found within a k-mer path. Since we don't know the exact number of nucleotide disagreements that may form multiple SNPs on each path, enumerating all the different sizes of cycles starting from a branching vertex can potentially return all the variants on the paths but at the expense of extended the path beyond $k+1$ steps.  This enumeration of cycles requires $k+n$, k is fixed, however $n$ takes values from 1 to $k$, since the paths of these bubbles expand by the number of disagreeing nucleotides that forms $k+n$ steps in paths.  This cycle enumeration process could be an expensive task as two loops are required. The first loop for the branching vertices and the second loop for $n$, which could take up to $O(n^2)$ in complexity.   We extend the $2k+2$ bubbles search method to take advantage of the multi-cores parallel computing found in modern CPUs. The aim is to search for all cycle sizes which could improves the SNP detection and classification down-stream analysis task. We show that the multi-core and multi-node design for our graph decomposition search method improves significantly cycle enumeration running time for colored de Bruijn graph based assemblers.

In the first part of the paper, we introduce $\Pi$-cyc application and its different modules.  Sections 2, describes related work. In section three, we describe the algorithms and methods used for building $\Pi$-cyc application. Section four is the experimental evaluation of our cycles search method and the comparative results.  Section five is a discussion and we conclude in section six.   

\section{Related Work}

The bubble caller developed for cortex\_var uses a single thread to search complex de Bruijn graphs \cite{cortex_2012}, a serious limitation for very large and multiple assembled genomes analysis. In contrast, $\Pi$-cyc concurrently searches very large multi color graphs for $2(k+n) + 2$ cycles using graph decomposition then uses a heuristic method for SNP predictions based on path coverages.  Thus, reducing the overhead of a single thread cycles enumeration and improves significantly the CPU time. cortex\_var bubble search was not designed for multiple genomes. And hence,  we can not use it in the comparison section of the paper.  In contrast, DiscoSNP++ application is a population based approach for de novo variant calling where samples or genomes, also using de Bruijn graph assembly based on bloom filter, searches for bubbles~\cite{disco}. However, the algorithm used for bubble search miss a large number of SNPs as shown in \ref{sec:snpcallinganalysis}.  DiscoSNP++ uses a breadth-first search approach while cortex\_var bubbles search uses Dijkstra's algorithm.  $\Pi$-cyc multi-node and multi-core approach effectively enumerates $2(k+n) + 2$ graph cycles, also known as nested bubbles, leading to the discovery of potentially important variants in reference-free sequenced samples.  

\section{The Method}

\begin{figure}[h!]
\centerline{\includegraphics[width=8cm, height=7cm]{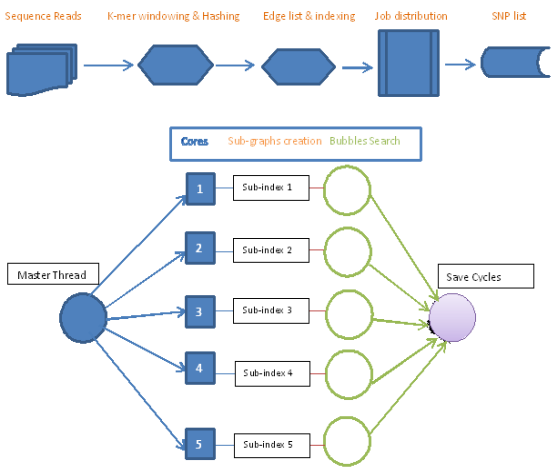}}
\caption{Algorithms and Work distribution in $\Pi$-cyc.}\label{fig:01}
\end{figure}

$\Pi$-cyc finds cycles by graph decomposition. The method is similar in principle to the method shown in~\cite{younsi2015using}. First, an edge list representing the linked de Bruijn graph is extracted from a binary colored de Bruijn graph. Second, undirected subgraph are created using the edge list for cycles enumeration.  An important difference is the extraction of the hash table index and the ad-hoc subgraph creation. A full multicolour de Bruijn graph stored in a hash table is indexed by its branching vertices.  This branching vertices index is divided among several compute nodes with multi-core CPUs to build undirected subgraphs of specific sizes.  Each CPU-core works with a sub-index for subgraphs creation and cycles enumeration.  The two processes are executed using the sub-indexes in parallel.  The main objective is to search, when possible, the entire de bruijn graph for specific cycles.  These cycles are shown to contain, with high probability, SNPs using their k-mer coverages \cite{younsi2015using}.  Figure \ref{fig:01} shows the application flow diagram that describes the main design.   

\subsection{Software main features}

The core part of $\Pi$-cyc assembly process, which uses cortex\_con assembler, contains two main subparts, windowing and hashing. The different algorithms that are combined to build $\Pi$-cyc application are:

\begin{enumerate}

\item Read Input files: Illumina reads from input files \footnote{We used cortex\_con C implementation http://cortexassembler.sourceforge.net/} 
\item Build a hash table: using K-mers $k-mer$ windowing and Hashing \footnote{See footnote 2}
\item Combined de Bruijn graphs is stored in a single Hash table
\item An index is calculated from the Hash table
\item The index is divided among CPU-cores  (see Algorithm~\ref{alg1})
\item Subgraphs are dynamically created using a fast neighborhood building block method in paralell (see Algorithm~\ref{alg2})
\item Searching for cycles is executed in paralell (see Algorithm~\ref{alg3})
\end{enumerate}

In general, the application uses several steps to build colored de Bruijn graph which is stored in a hash table, undirected subgraphs are created in parallel for cycles search. A full description of Cortex\_con can be found in \cite{cortex}.  

\begin{algorithm}
\SetKwData{Left}{left}
\SetKwData{This}{this}
\SetKwData{Up}{up}
\SetKwFunction{Union}{Union}
\SetKwFunction{GraphNeighborhood}{GraphNeighborhood}
\SetKwFunction{SearchCyc}{SearchCyc}
\SetKwInOut{Input}{input}
\SetKwInOut{Output}{output}
\Input{A finite Index array $I=\{a_1, a_2, \ldots, a_n\}$ of integers}
\Output{A set of Cycles}
\BlankLine
Initialise branching vector $b$\;
Initialise cycle set $s$\;
Initialise subgraph vector $g$\;
Set Kmer value $k$ \; 
Set minG = ((k * 2) + 2) (\tcp*[h]{smallest subgraphs size })\;
\For(\tcp*[h]{first openMP loop}){$i \gets 1$ \textbf{to} $n$ }{
$j\leftarrow  a_i$\;
$g\leftarrow$ \GraphNeighborhood{j}\;
$b\leftarrow  getbranchingvertice(g)$\;
\For(\tcp*[h]{second openMP loop}){$i\leftarrow 1$ \KwTo $b$}{
\If{b > 1 {\bf and} g > minG}{\label{lt}
\For(\tcp*[h]{third openMP loop}){$i\leftarrow 1$ \KwTo $(((k + i) * 2) + 2)$}{
$s\leftarrow$ \SearchCyc{$g$}\;
}

}
}
}
\caption{Parallel Subgraph creation and cycle search.}
  \label{alg1}
\end{algorithm}

Algorithm~\ref{alg1} describes the main loops used for subgraph creation and cycles enumeration. The first loop takes the computed index, or part of it in an HPC distributed system, and divide it among the CPUs.  Each CPU-core works with a part of the index to create subgraphs and search for cycles.  

The two main Algorithms~\ref{alg2}, used in subgraph creation, and~\ref{alg3}, used for cycle search, are:

\begin{itemize}
\item Build a subgraph using a fast neighborhood building block method.
\item Parallel $2(k+n) + 2$ cycles enumeration
\end{itemize}

The subgraph creation process described in Algorithm~\ref{alg2} uses a neighborhood building block method.  Initially, a branching vertex is selected from the subindex.  Every other adjacent vertex is selected and including in the subgraph creation and the process incrementally adds vertices to the size of the subgraph up to a user defined threshold.  Genome assemblies from multi-colored de Bruijn graphs can be very complex. The objective of a neighborhood method is to build subgraphs small enough to reduce graph complexity but large enough to search for $2(k+n) + 2$ cycles.    

\begin{algorithm}
\SetKwFunction{GraphEdgeList}{GraphEdgeList}
 \KwData{Branching vertex j from index}
 \KwResult{Subgraph g}
Set max subgraph size $v=x$, $x=minG,\ldots,n$; $n \in \mathbb{N} $. \;
Initialise subgraph vector $e=0$\;
Initialise edge list map $m=0$ and $i=0$\;
$e_0 \leftarrow  j$\;
 \While{ e $<$ v }{
  $m = GraphEdgeList(e_i)$(\tcp*[h]{get adjacent vertex $a_i$ from the hash table})\;
  \If{$a_i \notin m$}{
    $e \leftarrow  a_i$\;
    $i++$
   }
 }
 $g \leftarrow  m$\;
 \caption{Graph Neighbourhood.}
  \label{alg2}
\end{algorithm}

The cycle enumeration Algorithm~\ref{alg3} takes a branching vertex from the subgraph and walks $k + n + 1$ steps and records whether another branching vertex is found.  A full cycle is found when the walk returns to the starting vertex in $2(k+n)+2$ steps.  Only cycles that have equidistant branching vertices are kept for further analysis. Cycles are decomposed into paths and labels are read from the hash table with their potential SNPs.  The SNP prediction algorithm uses the extracted k-mer path coverages for each colour.

\begin{algorithm}
\SetKwFunction{SearchCyc}{SearchCyc}
 \KwData{Subgraph g}
 \KwResult{A Set S = $2(k+n) + 2$ cycles }
Set path Array p\;
Initialise cycle vector v\;
\For(\tcp*[h]{Select branching vertex b from g})
{$b \gets 1$ \textbf{to} $n$ }{
$p_i \leftarrow  b$\;
 \While{ p $<$ $2(k+n) + 2$ }{
  \eIf{$v_i \in v$}{
    $v = walk\_forward(g)$\;
   }{
    $v = walk\_reverse(g)$\;
  }
  }
 }
 $S \leftarrow  v$\;
  \caption{Walk the graph.}
   \label{alg3}
\end{algorithm}


\section{Results}

The Pombe whole genome DNA sequencing datasets (DNA-seq)~\cite{pombe2015} used for the CPU time evaluation and Cycle enumeration are described in Table~\ref{fig2}. 




\subsection{CPU Time Analysis}

We run the CPU time tests on a Dell EMC PowerEdge C6420 server~\footnote{https://wiki.helsinki.fi/display/it4sci/Kale+User+Guide} with a twin Skylake 20-cores hyperthreaded CPUs for a total of 2x40 cores.  The application was compiled with AVX2 option using intel C++ version 18.0.2 compiler (AVX-512 intel compiler option showed similar overall results).  

\begin{table}
  \begin{center}
    \caption{$2(k+n) + 2$ cycles found for different CPU-cores run for 60 minutes CPU time using the 10 Pombe genome strains in Table~\ref{fig2}.}
    \label{tab:table1}
\begin{tabular}{l|l|l|l}
\hline
cores&	total cycles	& used index	&time (s)\\
\hline
1&	669&	64&	3607.50\\
5&	2347&	212&	3648.70\\
10&	3886&	360&	3652.03\\
20&	7150&	680&	3604.22\\
30&	10583&	911&	3617.55\\
40&	13875&	1344&	4324.67\\
80&	19255&	1855&	3803.10\\
\hline
    \end{tabular}
  \end{center}
\end{table}

The results shown in Table~\ref{tab:table1} are from a combination of 10 Pombe genomes, as shown in Table~\ref{tab:datasets}.  The computed hash table index is 1314786 vertices and is divided among 5 compute nodes with 262957 vertices sub-index for each node. Each experiment is run with varying CPU-cores as shown in the table and with the same set of parameters. K-mer value $K = 63$ and the subgraph size is set to 5000 vertices.  The total number of cycles found for a single node is also shown in \textit{total cycles} column (similar figures are found on each node).  The \textit{used index} on Table~\ref{tab:table1} shows the total number of vertices in the sub-index used to create subgraphs.  The \textit{time} columns shows the total CPU time it took for the graph creation and cycle search after the allocated 60 minutes CPU time has ended.  The extra time accounts for time spent in the loops before it exited.

\begin{figure}
    \includegraphics[width=\linewidth]{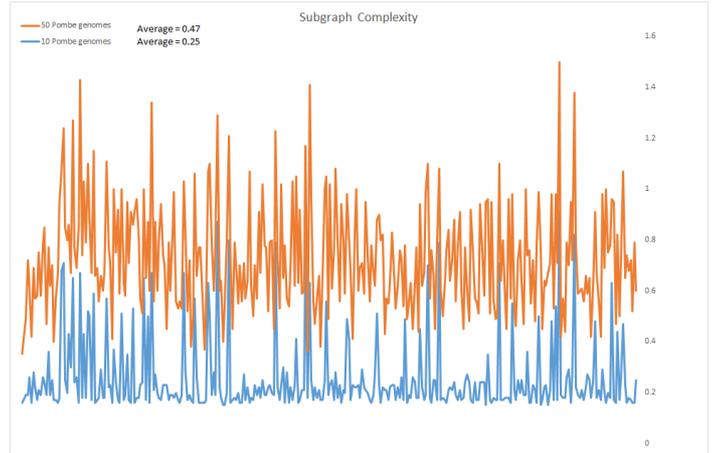}

    \caption{Graph complexity increases with the increase of the number of genomes.}
    \label{fig1}
\end{figure}

The CPU time of the graph creation and cycles search on a single core is an average of 669 $2(k+n) + 2$ cycles.  A substantial increase in shown for the 80 cores with 19255 cycles found for a total of 3803.1 seconds. It took an extra 3.38 minutes for the search to finish for the 80 cores.  The average CPU time of the 5, 10, 20, 30 and 40 cores parallel search is shown in Table~\ref{tab:table1}.  A parallel multi-core search that finds a similar number of cycles as found by 1 core will take far less time.  The decrease in CPU time is mostly noticeable for the 80 cores (i.e. a complete single compute node in our HPC server) which finds 4358 cycles in 811.78 seconds, about 6.5 folds increase in cycles found, and 4 folds decrease in CPU time in comparison to one CPU-core.  

The graph complexity is shown in figure~\ref{fig1}. A simple measure is to calculate the number of connected vertices in a given subgraph (or total edges) then divide by the total number of vertices in the subgraph for normalisation. We notice that the increase in subgraph complexity increases with the number of genomes used in the experiments. The average complexity measure for 10 Pombe genomes is 0.25 while for the 50 Pombe genomes is 0.47.  Similar samples that produces complex interconnected graphs will benefit from the parallel search method we developed.   

\begin{figure}
    \includegraphics[width=\linewidth]{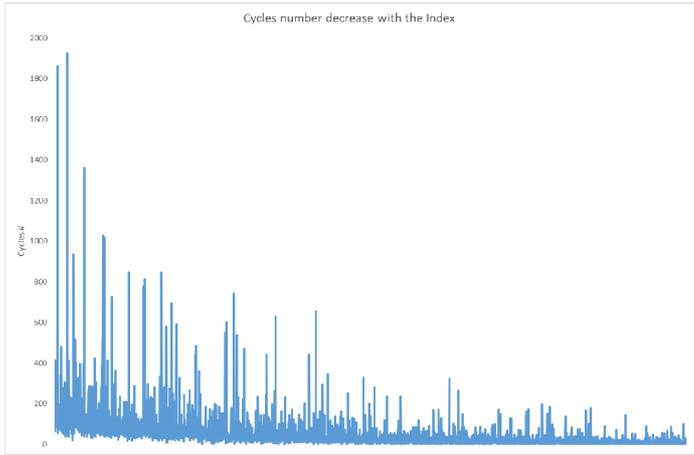}

    \caption{less new cycles are found after half of the index is used.}
    \label{fig2}
\end{figure}

In heavily interconnect graphs found in a multi-coloured setup, we noticed that as the search progress less cycles are discovered which eventually returns small number of cycles towards the end of the subgraphs creations as shown in Figure~\ref{fig2}.  Using the entire branching index in a heavily interconnect de Bruijn graph will result is searching for duplicate cycles and will require increased resources for searching the entire de Bruijn graph.  An efficient way to reduce the search load is to select only a third of the entire index for subgraph creation.


\subsection{SNP Calling Analysis}
\label{sec:snpcallinganalysis}

In order to assess the results of the contigs in the fasta file produced by the application, we use the blast program with default parameters to align each path to the reference genome \textit{Schizosaccharomyces Pombe ASM294v2}.  The SNP results are processed by an in-house program that search for nucleotide differences in the blasted file that match those in the produced contig fasta file. 

\begin{table}
  \begin{center}
    \caption{\textit{Schizosaccharomyces Pombe Strains}, 30-gen = 10-gen strains + the datasets, 50-gen = 30-gen strains + the datasets. 57-set shows the datasets used in the SNP calling method of~\cite{pombe2015}.}
    \label{tab:datasets}
\begin{tabular}{l|l|l|l|l|l}
\hline
\smaller{10-gen Strain}&	\smaller{57-set}	& \smaller{30-gen Strain} & \smaller{57-set}	& \smaller{50-gen Strain} & \smaller{57-set}\\
\hline
\smaller{ERR107690} & \smaller{JB4} & \smaller{ERR107700} & & \smaller{ERR107720} & \smaller{JB858} \\
\smaller{ERR107691} & \smaller{JB22} & \smaller{ERR107701} &	\smaller{JB837} & \smaller{ERR107721} & \\
\smaller{ERR107692} &	& \smaller{ERR107702}	& \smaller{JB838} &  \smaller{ERR107722} & \\
\smaller{ERR107693} &	& \smaller{ERR107703} & &  \smaller{ERR107723} & \\
\smaller{ERR107694} &	& \smaller{ERR107704} & \smaller{JB840} & \smaller{ERR107724}	& \smaller{JB862} \\
\smaller{ERR107695} &	& \smaller{ERR107705}	& \smaller{JB841} & \smaller{ERR107725} &	\smaller{JB864} \\
\smaller{ERR107696} &	& \smaller{ERR107706}	& \smaller{JB842} &\smaller{ERR107726} & \\
\smaller{ERR107697} &	\smaller{JB758} & \smaller{ERR107707} &	\smaller{JB845} & \smaller{ERR107727} & \\
\smaller{ERR107698} &	& \smaller{ERR107708} &	\smaller{JB846} &  \smaller{ERR107728}	& \smaller{JB869}  \\
\smaller{ERR107699} &	\smaller{JB762} & \smaller{ERR107709} & & \smaller{ERR107729}	 & \smaller{JB870}  \\
          &       & \smaller{ERR107711} & & \smaller{ERR107730}	& \smaller{JB873}  \\
          &       & \smaller{ERR107712} & & \smaller{ERR107731}	 &\smaller{JB875}  \\
          &       & \smaller{ERR107713} & & \smaller{ERR107732} & \\
          &       & \smaller{ERR107714} & \smaller{JB852}& \smaller{ERR107733} &	\smaller{JB878}  \\
          &       & \smaller{ERR107715} & \smaller{JB853}&	\smaller{ERR107734}&	\smaller{JB879}  \\
          &       & \smaller{ERR107716} & \smaller{JB854}& \smaller{ERR107736}&	\smaller{JB884}  \\
          &       & \smaller{ERR107717} & &	\smaller{ERR107737}  & \\
          &       & \smaller{ERR107718} & &	\smaller{ERR107738}  & \\
          &       & \smaller{ERR107719} & &	\smaller{ERR107739}  & \\
\hline
    \end{tabular}
  \end{center}
\end{table}

We made certain to select randomly genomes from 57 strains included in~\cite{pombe2015} SNP calling analysis. A summary of the Table~\ref{fig2} includes 3 of the 57 datasets used in \cite{pombe2015} in the 10-Pombe experiments, 13 and 25 for the 30-pombe and 50-pombe experiments.

Table~\ref{table2} shows the results of the Pombe datasets, using $k = 63$ for the de Bruijn graph construction, for a single compute node. The application generated approximately 200000 cycles for a combination of 10 genomes, 500000 for 30 genomes and a million for 50 genomes. Filtering these sets keeps only cycles that have relatively matched paths using a parameter allowing maximum mismatch value $f = 15$ and shown in \textit{Fil} column.  The number of predicted SNPs using path coverages are shown in \textit{Pred} column.  We used a SNP validation set as reported in~\cite{pombe2015} to confirm the SNP found in the filtered sets and reported as \textit{Matched} in Table~\ref{table2}.  The \textit{Index} column shows the branching vertices of the coloured de Bruijn graph.  The \textit{subG} column shows the total number of subgraphs processed using a maximum of 5000 vertices per subgraph.  The index values was divided among 10 compute nodes for the 10 and 50 genomes experiments and 15 nodes for the 30 genomes experiment.

\begin{table}[ht]
  \begin{center}
    \caption{$\Pi$-cyc Results using 10, 30 and 50 Pombe genome strains. \#: number of genomes used. Cyc: number of cycles found. Fil: using parameter $f = 15$ for maximum allowed number of mismatches in paths. Pred: \# of SNP predicted using path $k-mer$ coverages. Matched: SNPs that matched the the variant set reported in~\cite{pombe2015}. Index: the total index value calculated from the hash table.  SubG: the number of subgraphs created using the index value.}
    \label{table2}
\begin{tabular}{l|l|l|l|l|l|l}
\hline
\#& Cyc & Fil & Pred & Matched & Index & SubG \\
\hline
10  & 209067  & 154406 & 91008 & 84806 & 1314786 & 752760 \\
30 	& 547096  & 206060 & 120396 & 110410 & 3395397 & 1251465 \\ 
50 	& 1055245 & 258366 & 157127  & 122631 & 5086954  & 897600 \\		
\hline
    \end{tabular}
  \end{center}
\end{table}

Many of the cycles in the \textit{Cyc} row shown in Table~\ref{table2} are indels that are not reported in this paper.  The 10-genomes experiments found 84806 SNPs, 30-genomes experiment 110410 SNPs, and the 50-genomes experiment found 122631 SNPs that matched the SNP set. The increase in number of SNPs found with the increase in the number of genomes used, and the relatively small size of index used for subgraph creation indicates the design approach we used is useful to enumerate a large number of SNPs when combining many genomes. The graph complexity is indeed a major issue for very large number of colours.  

\begin{table}
  \begin{center}
    \caption{DiscoSNP++ Results Analysis.  \#: number of genomes used. Matched: SNPs that matched the SNP set. bwa\_coh: Found bubbles in the coherent set are run with bwa on the reference genome. bwa\_uncoh: Found bubbles in the uncoherent set are run with bwa on the reference genome.}
    \label{tab:comp}
\begin{tabular}{l|l|l|l|l|l}
\hline
\#  & bwa\_coh & Matched  & bwa\_uncoh & Matched & tot\\
\hline
10 & 57006 & 53347 & 12710 & 10876 & 64223 \\ 
30	& 90740 & 82926 & 15169 & 11118 & 94044 \\ 
50	& 103424 & 99110 & 17732 & 12722 & 111832  \\		
\hline
    \end{tabular}
  \end{center}
\end{table}

Table~\ref{tab:comp} compares our method to DicoSNP++ application, a similar de novo multi-colour de Bruijn graph bubble search that uses Bloom filter ~\cite{disco}. We run the experiments without a reference genome to avoid any bias in bubble search.  The parameters used for DiscoSNP++ are $k=63$, b=2, P=15, and the remaining are kept as default values. According to the user document that comes with the application, b and P are the two parameters that are directly involved in bubble search. We choose $b=2$ as we wanted to get as many bubbles as possible from DiscoSNP++ results in order to align the paths with the Pombe reference genome, and $P=15$ to allow for large number of SNPs to be reported per bubble. The default values are $b=1$ and $P=1$, and as reported by the authors will ensure a high accuracy but low recall. Our target is to enumerate all bubbles regardless of their accuracy then match them to the canonical SNP set.  The command used to run DiscoSnp++:\\

run\_discoSnp++.sh -r genomesfastq.txt -k 63 -t -b 2 -P 15 -g -G Schizosaccharomyces.fa -B /bwa/
\\
\\
The results shown in Table~\ref{tab:comp}, demonstrate that our method finds more SNPs in comparison with DiscoSNP++, with 24\% increase for the 10 genomes experiment, 15\% increase on 30 genomes, and 8.8\% for the 50 genomes.  The 30-genomes and 50-genomes experiments used a small number of subgraphs in comparison to the maximum index value.  We expect more SNPs discovery as more subgraphs are searched.  Although, in our opinion, searching half of the total index is sufficient to enumerate a large number of cycles as shown in the graph of Figure~\ref{fig2}.



\begin{figure}
    \includegraphics[width=0.50\linewidth]{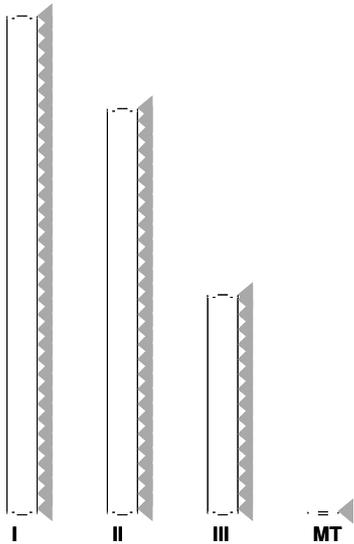}

    \caption{SNPs distribution 50 Pombe genome SNPs results as found by  $\Pi$-cyc}
    \label{fig3}

\end{figure}

\begin{figure}
    \includegraphics[width=\linewidth]{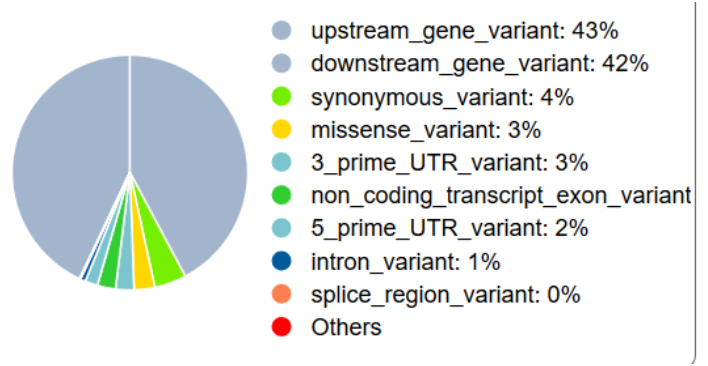}

    \caption{Variant Effect Predictor Analysis of 50 Pombe genome SNPs}
    \label{fig4}

\end{figure}

Figure~\ref{fig3} shows the distribution of SNP found by the 50-genomes experiment.  Our methods finds SNP in all the regions of the genomes without bias, and Figure \ref{fig4} shows the SNP consequences results on genes for the 50-genomes experiment matched SNP set using the Variant Effect Predictor tool~\cite{McLaren2016}.

\section{Discussion}


The multi-genomes coloured graph produced by the raw assembly can be big, yet simple in structure and complexity if raw sequencing reads are cleaner in comparison to smaller graphs with very noisy reads.  Cycles enumeration takes far longer to run in complex interconnected graphs built with large nucleotides sequencing error rates. A parallel graph decomposition, implemented as a neighbourhood subgraph as shown in this paper, lowers graph vertices interconnection complexity while the distributed search using multiple compute nodes and CPU-cores decrease dramatically the overall cycle's search.  This type of graph decomposition is also ideal for a hybrid CPU/GPU compute node.  Running large complex genomes simultaneously will have a major speed boost in de novo coloured de Bruijn graph for variants calling.

We used Cortex\_con de Bruijn graph assembler which is outdated and it hasn't been checked for code optimisation.  We believe our parallel method can be used with any couloured de Bruin graph assembler as part of a variant analysis pipeline.  In general, algorithms manipulating data externally are well-know to be slower in comparison to full memory based representation, we believe that our graph decomposition method can be used successfully for very large number of colours using DBG. In addition, we believe that a hybrid GPU/CPU platform could be used efficiently to accelerate the search phase in very complex multi-colour de Bruijn graph.  Graph error cleaning can improve drastically genome assembly and our method can be used in this context with no modification. 

The other main feature of the application is the down stream data analysis program for SNP predictions. The coverage pattern exhibited in cycles can be classified as either an allele in a heterozygous genome, a SNP or a repeat.  However, genome sequences are not free from errors which is the reason path coverages can be misleading.  Machine learning can deal with samples that contain noise and outliers which make it an ideal field to analyse coverage datasets ~\cite{younsi2015using}. We developed a simple coverages method using a heuristic path analysis approach and we are looking at ways in extending the machine learning classification approach first proposed in ~\cite{younsi2015using}.

The main issue encountered while designing $\Pi$-cyc is the trade-off between CPU speed and memory overhead.  That is, the additional time to access memory and the extra space needed for graph storage. An MPI architecture distributed among several compute nodes incurs a higher time in memory access while in a shared memory architecture, using C++ openMP as an example, the issue is overcome.  Especially that HPC compute node are increasing their CPU-cores and memory dramatically. However, the actual implementation uses stored hash table files that are uploaded on each compute node memory to increase CPUs availability for the parallel graph search.  An alternative approach is using message passing interface on a single node to send requested edge list for subgraph creation. In this case, the speed and memory trade-off is also a design issue to be taken into consideration. 

Furthermore, we believe that our parallel search method can be incorporated as a plugin for the several emerging de novo based colored de Bruijn graph assemblers using succinct data representation (SDBG).  Succinct de Bruijn graphs showed promising results in compression using very large number of colours.  SDBG have indeed shown that coloured graphs can be used in de novo population based genomics assembly for variant calling especially that modern sequencing technologies are increasingly producing cleaner samples and in large quantities. The emphasis is on developing cycle enumeration algorithms for variant calling that are fast and efficient. Our "all versus all" method would be suitable for comparing a very large number of genomes simultaneously with or without a reference genome. We showed that a single HPC compute node with $80$ CPU-cores can be used efficiently with $\Pi$-cyc application for SNP calling in transformed de Bruijn graph. We believe that our method is a step forward for using heterogeneous platforms, such as those found in the HPC compute nodes.

\section{Conclusion}

In this paper, we reported a novel design to parallelise the computational intensive task of finding bubbles in coloured de Bruijn graphs. The parallelism we introduced is incorporated in the graph creation and cycles search.  The graph decomposition proposed take advantage of multi-core CPUs found in multi-node platforms.  Our main approach is to allocate a portion of the graph index to all available multi-core CPUs in a shared memory compute node, and also designed to be use in a multi-nodes in a distributed memory platform.  Cycle enumerations algorithms in big and complex graphs are time consuming.  We showed that using our method, on an HPC parallel platform, can help reduce considerably the CPU time of a single/dual CPU-core machine.  The time required to search complex graphs for specific sizes cycles depends on the number of machines available to the user. Found cycles can then be translated into SNP using path coverages.   Our overall application can help the scientific community work with large and complex de novo genomes for SNP calling.  We identified a number of possible parallel extensions that will improve graph building and memory access. In addition, we believe using heterogeneous GPU/CPU platforms can increase dramatically the acceleration of graph search which will allow the analysis of extremely large number of big genomes simultaneously, thus benefiting population and comparative genomic research. 

\section*{Acknowledgements}
The programming and testing phases were used on Taito HPC servers at the Finnish IT center for science in Espoo (CSC Finland: https://www.csc.fi/).
Our experiments were conducted on Kale and Ukko2 HPC servers located at the University of Helsinki (https://wiki.helsinki.fi/display/it4sci/)

\section*{Funding}

Conception, implementation and testing by RY. This work was supported in part by JT with the Academy of Finland grant 317680. 


\bibliographystyle{natbib}

\bibliography{document}

\end{document}